# One-ninth magnetization plateau stabilized by spin entanglement in a kagome antiferromagnet

Sungmin Jeon,[1,12] Dirk Wulferding,[2,3,12] Youngsu Choi,[4] Seungyeol Lee,[1] Kiwan Nam,[5] Kee Hoon Kim,[5] Minseong Lee,[6] Tae-Hwan Jang,[7] Jae-Hoon Park,[7,8] Suheon Lee,[9] Sungkyun Choi,[9] Chanhyeon Lee,[10] Hiroyuki Nojiri,[11] and Kwang-Yong Choi[1*]

[1]*Department of Physics, Sungkyunkwan University, Suwon 16419, Republic of Korea*

[2]*Center for Correlated Electron Systems, Institute for Basic Science, Seoul 08826, Republic of Korea*

[3]*Department of Physics and Astronomy, Seoul National University, Seoul 08826, Republic of Korea*

[4]*Department of Energy Science, Sungkyunkwan University, Suwon 16419, Republic of Korea*

[5]*Center for Novel States of Complex Materials Research, Department of Physics and Astronomy, Seoul National University, Seoul, 08826, Republic of Korea*

[6]*National High Magnetic Field Laboratory (NHMFL), Los Alamos National Laboratory (LANL), Los Alamos, New Mexico 87545, USA*

[7]*MPPHC-CPM, Max Planck POSTECH/Korea Research Initiative, Pohang 37673, Republic of Korea*

[8]*Department of Physics, Pohang University of Science and Technology, 37673, Republic of Korea*

[9]*Center for Integrated Nanostructure Physics, Institute for Basic Science, Suwon 16419, Republic of Korea*

[10]*Department of Physics, Chung-Ang University, Seoul 06947, Republic of Korea*

[11] *Institute for Materials Research, Tohoku University, Katahira 2-1-1, Sendai 980-8577, Japan*

[12] These authors contribute equally: Sungmin Jeon and Dirk Wulferding

[*] Email to choisky99@skku.edu

**The spin-1/2 antiferromagnetic Heisenberg model on a Kagomé lattice is geometrically frustrated, which is expected to promote the formation of many-body quantum entangled states. The most sought-after among these is the quantum spin liquid phase, but magnetic analogs of liquid, solid, and supersolid phases may also occur, producing fractional plateaus in the magnetization. Here, we investigate the experimental realization of these predicted phases in the Kagomé material YCu$_3$(OD)$_{6+x}$Br$_{3-x}$ ($x \approx 0.5$). By combining**

**thermodynamic and Raman spectroscopic techniques, we provide evidence for fractionalized spinon excitations and observe the emergence of a 1/9 magnetization plateau. These observations establish $YCu_3(OD)_{6+x}Br_{3-x}$ as a model material for exploring the 1/9 plateau phase.**

**Introduction**

One of the most sought-after many-body phenomena in the field of quantum magnets is a quantum spin liquid (QSL)- a long-range entangled spin state that negates the standard symmetry-breaking paradigm. QSLs feature fractionalized quasiparticles and emergent gauges[1-4]. In the quest to realize QSLs, an $s=1/2$ Kagomé Heisenberg antiferromagnet (KHAF)- a two-dimensional lattice formed by corner-sharing triangles- is proposed for a paradigmatic setting for emergent QSLs. Accumulating numerical evidence supports the existence of a QSL ground state in the KHAF, yet its precise nature remains unsettled, with two potential flavors: a $Z_2$ gapped QSL and a gapless $U(1)$ liquid[5-13].

Aside from the likely zero-field QSL, the application of a magnetic field to the KHAF engenders exotic quantum phases at fractional magnetization plateaus $m=1/9, 3/9, 5/9$, and $7/9$[14-28]. The kinetic frustration induces the crystallization of magnons localized on the hexagon of the Kagomé lattice at $m=3/9, 5/9$, and $7/9$, resulting in spin-gapped solid phases. Additionally, a supersolid phase emerges just below the $m=5/9$ plateau. On the other hand, the nature of the 1/9 magnetization plateau, which is stabilized by spin entanglement, remains subject to controversy, with the possibility of a topological $Z_3$ spin liquid vs. valence bond crystals[21,26,28]. Despite the fundamental interest in the 1/9 plateau phase, its experimental realization has proven elusive due to the lack of adequate materials, presenting a longstanding challenge in Kagomé physics.

To date, a number of $s=1/2$ Kagomé compounds, including herbertsmithite $ZnCu_3(OH)_6Cl_2$, kapellasite $\alpha$-$Cu_3Zn(OH)_6Cl_2$, and Zn-barlowite $Cu_3Zn(OH)_6FBr$[29-36], have been reported as QSL candidates. These Kagomé materials commonly bear key QSL characteristics, such as fractionalized spin excitations and the absence of long-range magnetic order. Notwithstanding, the presence of Zn-Cu antisite mixing hinders the determination of the intrinsic ground state of KHAFs. Rather, this mixing along with lattice imperfections and magnetic disorder leads to the coexistence of gapped singlets and gapless states[36], masking the pristine Kagomé QSL state.

The recently discovered YCu$_3$(OD)$_{6+x}$Br$_{3-x}$ (abbreviated as YCu$_3$) is a promising addition to the Kagomé QSL system[37-39]. Unlike herbertsmithite, YCu$_3$ is nearly free from site-disorder-induced orphan spins, while the random distribution of OD$^-$/Br$^-$ introduces exchange randomness. Despite this quenched bond randomness, a Kagomé lattice of Cu$^{2+}$ is globally undistorted. Singularly, YCu$_3$ exhibits no magnetic ordering or freezing down to 50 mK and its specific heat displays a nearly quadratic $T$ dependence, reminiscent of a Dirac-like QSL. Considering a moderate exchange interaction of $J$~63 K (see below) and negligible intersite mixing, YCu$_3$ can serve as a valuable model system for investigating the predicted 1/9 plateau phase.

Here we present a comprehensive study of the zero-field and field-induced phases in YCu$_3$ by employing various thermodynamic and Raman spectroscopic techniques. Our specific heat, thermal conductivity, and Raman scattering data evince the presence of fractionalized spinon excitations. The most salient finding is the observation of the highly sought-after 1/9 plateau phase at a moderate field of $\mu_0 H_{1/9}$=15-21 T, making it accessible for further experimental exploration.

**Results**

**Specific heat and thermal conductivity** To gain insights into the ground state and low-lying excitations, we first turn to the specific heat and thermal conductivity of YCu$_3$. Figure 1a presents the $T$ dependence of the magnetic specific heat $C_m(T)$, obtained by subtracting a phonon contribution from the total specific heat (see Supplementary Fig. 1). Our $C_m(T)$ data match closely with numerical calculations for a KHAF[39] (pink line and green triangles), capturing the high-$T$ broad maximum at $T_{max}$~0.67$J$. From the maximum temperature $T_{max}$=42 K, we estimate the exchange interaction as $J$~63 K. On heating toward $T$~4$J$, the magnetic entropy $S_m(T)$, calculated by integrating $C_m/T$, approaches the expected value $R\ln 2$ for an $s$=1/2 system (dashed line in Fig. 1b). Compared to the theoretical curves, the low-$T$ behavior of $C_m$ reveals two anomalies: a suppressed hump feature at $T$~0.1$J$ and a weak peak at $T^*$~2 K that are ascribed to bond randomness and Dzyaloshinskii-Moriya (DM) interaction born from bond randomness[40], respectively. However, only a small amount of entropy (3% of $R\ln 2$) is released below the $T^*$~2 K anomaly, indicative of the freezing of a small fraction of local magnetic moments. By resorting to the $T^*/D_c$≈0.91 relation[40] between the $T^*$ anomaly in the specific heat and the out-of-plane DM anisotropy $D_c$, we further determine $D_c$≈2.2 K (=0.035$J$), which is significantly smaller than the critical value of $D_{crit}$~0.1 $J$ that separates a QSL from a magnetically ordered phase[41]. We stress that unlike in other Cu-based Kagomé compounds, the out-of-plane DM anisotropy in YCu$_3$ is not strong enough to disrupt the QSL phase. At the very low temperatures of $T$=0.3-1 K, we observe a quadratic $T$ dependence $C_m \propto T^2$. Below 0.3 K, deviations from the $T^2$ behavior is ascribed to the nuclear Schottky contribution or destabilization of a QSL against random singlets[42].

Thermal conductivity $\kappa_{xx}$ conveys valuable information about the presence or absence of itinerant low-energy excitations. Figure 1d plots $\kappa(T)$ vs $T$ on a log-log scale. Notably, $\kappa_{xx}(T)$ shows a gradual increase with increasing $T$ across the whole measured temperature range. This behavior suggests that phonons alone are insufficient to account for the heat transport. Furthermore, the semilog plot of $\kappa_{xx}/T$ vs $T$ (Fig. 1e) reveals distinct humps and a dip in the $\kappa$ vs $T$ data. The dip feature of $\kappa_{xx}(T)$ alludes to the suppression of fractionalized excitations, possibly due to enhanced phonon-spinon scattering. To inspect whether the low-$T$ $\kappa_{xx}$ data contain signatures of itinerant spinons, we analyze the $\kappa_{xx}/T$ vs $T$ plot in Fig. 1f by fitting it to $\kappa/T = \kappa_0 + aT$. We identify a quadratic $T$ dependence below 0.6 K and a negligibly small $\kappa_0$, consistent with recent thermal conductivity measurements[43]. The latter suggests that itinerant fermionic contributions are largely quenched due to the bond randomness. Overall, our $\kappa_{xx}$ data signify the impact of phonon-spinon interactions and bond randomness on the heat conduction of the YCu$_3$ system.

**Spinon-antispinon pair excitations** To elucidate the characteristics and dynamics of fractional spinon excitations, we conducted Raman scattering experiments in (*XX*) and (*XY*) polarization configurations, corresponding to the $A_{1g}+E_g$ and $E_g$ symmetry channels, respectively. In Fig. 2 and Supplementary Figs. 2-6, we present the temperature and polarization dependencies of Raman spectra, which comprise sharp phonon peaks superimposed onto a broad magnetic continuum. Hereafter, we focus solely on the magnetic excitations obtained after subtracting the phonon peaks from the as-measured spectra. Further, we calculate the magnetic Raman susceptibility $\chi''(\omega)$ by correcting the Bose factor $n(\omega) = 1/[1 \exp\left(\frac{\hbar\omega}{k_B T}\right) - 1]$ using the relation $I(\omega) = (1 + n(\omega))\chi''(\omega)$. As plotted in Fig. 2a, the magnetic continuum centered around 15 meV at low temperatures experiences a gradual softening and damping with increasing $T$.

To a first approximation, the magnetic Raman spectral weight is dominated by spin-number-conserving scattering processes ($\Delta S = 0$). In the case of a KHAF, these processes involve one-pair (1P) and two-pair (2P) spinon-antispinon excitations[44]. In order to differentiate each contribution, we decompose the magnetic $\chi''(\omega)$ into three Gaussian profiles, denoted by the color shadings (P1, P2, and RS) in Fig. 2b,c. We discuss other models in Supplementary Figs. 4-5. The 1P excitation is centered at 13 meV ($\approx 4.6\chi J$) and the 2P component has a maximum at $T^{exp}_{2p}$=22 meV ($\approx 7.8\chi J$) and extends up to $T^{exp}_{2c}$=36 meV ($\approx 12.8\chi J$) with the adjustable parameter $\chi$=0.5($\approx 2\chi_s$). Here we note that the $\chi$ parameter is linked to the renormalization parameter $\chi_s$=0.221 of the spinon bandwidth[6,39]. In particular, the observed peak energy and spectral width of the 2P excitation are in accordance with the theoretical values of $T^{th}_{2p}$=(7-8)$\chi J$ and $T^{th}_{2c}$=12$\chi J$, which are computed for a Kagome QSL state. These experimental observations are consistent with the spectra reported in several spin-$\frac{1}{2}$ Kagomé systems, including herbertsmithite and Zn-Barlowite[44-47].

In contrast to YCu3, its magnetic counterpart YCu$_3$(OH)$_6$Cl$_3$ exhibits a sharp excitation in the magnetically ordered state below $T_N$=15 K[48], which corresponds to one-magnon scattering born out of the 1P excitation (refer to Supplementary Fig. 5). The lack of magnon excitations validates the spinon assignment. Apart from the prominent spinon excitations, we are able to identify another magnetic contribution (RS) at frequencies below 5-6 meV, which is attributed to bond-randomness-induced random singlets. Besides, a comparison of the $\chi''_{E_g}(\omega, T)$ and $\chi''_{A_{1g}}(\omega, T)$, shown in Fig. 2b,c, reveals their close resemblance, except for the dominance of the 2P component over the 1P component in the $A_{1g}$ spinon continuum. This trend is in compliance with the theoretical prediction[44]. However, the only partial suppression of the 1P spectral weight in the $A_{1g}$ channel may be due to the random occupation of OD$^-$/Br$^-$.

As evident from the color map of the Raman intensity $I_{E_g}(\omega)$ and $\chi''_{E_g}(\omega, T)$ in Fig. 2d,e, the spinon excitations dissolve into quasielastic scattering at the temperature scale of $J$=63 K (see the horizontal dashed lines). The 1P and 2P excitations are resolvable for temperatures below $J$, while their peak energies slightly increase as $T$ approaches 0 K (Supplementary Fig. 7-10). By employing the Kramers-Kronig relation, we derive the dynamic Raman susceptibility $\chi^{dyn} = \frac{2}{\pi}\int_0^\infty \frac{\chi''(\omega)}{\omega} d\omega$ from the magnetic Raman response. Figure 2f plots the temperature dependence of $\chi^{dyn}(T)$ evaluated by integrating the Raman conductivity $\chi''(\omega)/T$ up to 45 meV in the $A_{1g}$ and $E_g$ channels. For temperatures above $J$, $\chi^{dyn}_{E_g}(T)$ and $\chi^{dyn}_{A_{1g}}(T)$ scale perfectly with the static susceptibility $\chi_{dc}^{th}(T)$ computed for an ideal Kagomé antiferromagnet. Compared to $\chi_{dc}^{th}(T)$, however, $\chi_{E_g}^{dyn}(T)$ and $\chi_{A_{1g}}^{dyn}(T)$, defined as the dynamic limit of $\chi^{static} = \lim_{k\to 0}\chi(k,\omega = 0)$, is systematically reduced for temperatures below $J/2$ and $J$, respectively. This behavior is largely consistent with the repression of spinon excitation observed in the $A_{1g}$ channel[44] (Fig. 2b,c).

**Power-law dependence of spinon Raman susceptibility** Next, we examine the low-energy Raman susceptibility. We attempted to fit the magnetic Raman susceptibility to a power-law $\chi''(\omega) \propto \omega^\alpha$ in the energy range below 10 meV, in which the contribution from 2P spinon excitations are negligible. As shown in Fig. 3a-d, we could identify two distinct power-law behaviors in two energy intervals of (i) $\omega$=1- 5 meV and (ii) $\omega$=5- 10 meV for both scattering channels. The piece-wise power-law behavior stems from the random-singlet-like excitations present up to 5-6 meV (RS in Fig. 2b). The extracted exponents as a function of temperature are plotted in Fig. 3e, f. In the $\omega$=1- 5 meV regime, on cooling through $J$=63 K (yellow shading), the exponent drops from α=0.7 to 0.5 for the $E_g$ channel while it increases from α=0.15 to 0.72 for the $A_{1g}+E_g$ channel. We recall that the sublinear behavior is a hallmark of random singlets[42], which create an abundant low-energy DOS of the form $\rho(\omega)\sim\omega^{-\gamma}$. Thus, we conclude that the presence of bond randomness modifies the spectral form of the low-energy DOS, as hinted by the thermodynamic quantities discussed previously. In the $\omega$=5-10 meV interval, on approaching zero kelvin, the exponents increase to α=1.15 and 1.72 for the $E_g$ and $A_{1g}+E_g$ channels, respectively. This switch from a sublinear (α < 1) to a superlinear behavior (α > 1) in the magnetic $\chi''(\omega)$ suggests that the spinon excitations inherent to KHAFs prevail at higher temperatures, where perturbative effects become insignificant.

**1/9 magnetization plateau** Finally, we address the possibility that YCu$_3$ hosts a desired 1/9 magnetization plateau state. The $s$=1/2 Kagomé model predicts a series of magnetization plateaus at $m$=1/9, 1/3, 5/9, and 7/9. Among these plateaus, the 1/3, 5/9, and 7/9 plateaus can be envisioned as zero, one, and two hexagonal magnons while the edge spins of the hexagram are fully polarized[21-24]. Contrarily, the 1/9 plateau is rooted in a different origin involving entangled spins.

Figure 4a presents the high-field magnetization curve $m=M/M_S$ plotted against $\mu_0H/J_m$ ($J_m$=45 K *vide infra*) at $T$=2 K for the field orientation $H//ab$. The pulsed-field $M(H)$ (magenta circles) is scaled to the low-field SQUID data (yellow triangles) and normalized by the saturation magnetization $M_S=gS\mu_B$ ($g$=g-factor). As the applied field increases up to 60 T, YCu$_3$ exhibits two consecutive magnetic phases, inferred from a two-steplike increase of $m(H)$ at $\mu_0H_{1/9}\approx$18-23 T and $\mu_0H_{1/3}\geq$50 T. The dips in the field derivative d$m$/d$H$ (gray squares) confirm the presence of magnetization plateaus. To identify their origin, we compare the experimental $m(H)$ to the calculated magnetization process (blue line) for the spin-1/2 KHAF[26]. The lower-field magnetization plateau is associated with the predicted 1/9 plateau.

To investigate the stability of the 1/9 magnetization plateau, we further measured in-plane and out-of-plane magnetization up to 25 T at $T$=0.5 K. Our $^3$He magnetization data, presented in Fig. 4b, substantiate a robust $m$=1/9 plateau when the magnetic field is applied perpendicular to the Kagomé plane ($H//c$; azure squares). The lower and upper critical fields correspond to $\mu_0H^c_{c1}$=15.4 T and $\mu_0H^c_{c2}$=18.3 T, respectively (vertical arrows in the violet line). In the case of in-plane $m(H)$, the expected 1/9 plateau exhibits a finite slope, while the critical fields shift to higher fields $\mu_0H^{ab}_{c1}$=16.8 T and $\mu_0H^{ab}_{c2}$=21.1 T (vertical arrows in the coral line). The observed anisotropic behavior of $m(H)$ with respect to the field orientation is attributed to the $g$-factor anisotropy, DM anisotropy, and in-plane longer-range interactions. The magnitude of the out-of-plane DM component amounts to $D_c\approx$2.2 K, which is found to be comparable to the width of the out-of-plane $m$=1/9 plateau ($\Delta\mu_0H_{1/9}\approx$2.9 T). The robustness of the out-of-plane $m$=1/9 plateau suggests that perturbative terms, including the $D_c$ component, are not sufficiently strong to destabilize the $m$=1/9 state. Conversely, the finite slope observed in the in-plane $m(H)$ within the anticipated plateau regime indicates that the in-plane DM component $D_{ab}$ and in-plane longer-range interactions induce a mixing of the $m$=1/9 ground state with its excited states. In $s$=1/2KHAFs, it is known that the in-plane DM component $D_{ab}$ enhances the magnetic susceptibility[41]. Indeed, we observe the low-$T$ upturn of $\chi(T)$, as shown in Supplementary Fig. 1a. However, we note that the bond randomness can also contribute to the enhancement of $\chi(T)$ at low temperatures due to the rich low-energy DOS. Thus, further investigation is needed to accurately determine the strength of the $D_{ab}$ component and to elucidate whether its putative large value ($D_{ab} > D_c\approx$2.2 K) is primarily responsible for the instability of the $m$=1/9 phase.

Finally, by comparing the theoretical predictions with the out-of-plane $m(H,T$=0.5 K), the effective exchange interaction $J_m$ is evaluated as $J_m$=45 K, which is smaller than the estimated $J\sim$63 K by the specific heat measurements. This difference between $J_m$ and $J$ is attributed to the influence of a distribution of exchange interactions [$J$-$\Delta J$, $J$+$\Delta J$]. The magnitude of this exchange randomness can be estimated as $\Delta J=|J - J_m|\sim 18$ K, considering that $J$ is deduced from the centroid of $C_m$, while $J_m$ is determined by the onset of the $m=1/9$ magnetization plateau.

**Discussion and conclusion** Based on our experimental data, it is evident that YCu$_3$ occupies a unique parameter range among Kagomé antiferromagnets. Despite the presence of small DM anisotropies and moderate bond randomness, the predicted 1/9 magnetization plateau remains robust and immune to perturbative interactions when a magnetic field is applied perpendicular to the Kagomé plane. This may be owed to the fact that the out-of-plane DM component ($D_c \ll 0.1J$) alone is not strong enough to destroy a QSL ground state. On the other hand, the in-plane perturbative terms, including the in-plane $D_{ab}$ component and longer-range interactions, could be stronger than the out-of-plane terms, thereby leading to the destabilization of the $m=1/9$ plateau. While the field-induced phase exhibits some resilience against perturbations, the unavoidable perturbative interactions seem to modify the intrinsic Kagomé QSL. Given the significance of distinguishing between different QSL scenarios in the Kagomé lattice, we delve into the ramifications of our thermodynamic and spectroscopic data in relation to a $Z_2$ gapped versus a gapless $U(1)$ liquid state. The notion of dynamical crossover and persistent typicality in an incipient QSL provides a conceptual framework for the following discussion[49].

A key characteristic of QSLs is the emergence of fractionalized excitations, as demonstrated by our Raman scattering data (refer to Fig. 2 and 3). It is worth noting that the Raman spectrum and its associated susceptibility were specifically computed for the U(1) Dirac QSL, which provides a starting point for detailed comparison. In an ideal KHAF, the low-$\omega$ magnetic Raman spectrum follows $\omega^3$ dependence, as long as the matrix element vanishes for all 1P spinon-antispinon excitations[44]. However, when perturbative interactions introduce a finite matrix element, $\chi''(\omega)$ emulates the DOS of low-energy excitations, often leading to a weaker $\omega^\alpha$ dependence ($1<\alpha<3$). Since Dirac spinons possess a linear dispersion, $\alpha=1$ occupies a special point. In this context, the superlinear power-law behavior ($\alpha=1.15$ for the $E_g$ scattering channel and 1.72 for the $A_{1g}+E_g$ channel) observed in the energy range of $\omega=5-10$ meV suggests that the high-energy spinons retain a Dirac-like nodal structure. Below 5 meV, however, the YCu$_3$ system is substantially affected by perturbations, and the persistent typicality of Dirac spinons is no longer valid. Consequently, the lower-energy fractional excitations experience some modifications from the Dirac dispersion. This deviation is reflected in the sublinear power-law behavior observed in the Raman spectra below 5 meV.

Next, we turn to the thermodynamic and NMR signatures of QSLs. In the case of the Dirac

QSL, both the specific heat and the NMR spin-lattice relaxation rate $1/T_1$ are expected to exhibit a $T^2$ dependence[6]. Indeed, we observe a $C_m \propto T^2$ behavior over a limited temperature range of $T$=0.3-1 K. However, the NMR spin-lattice relaxation rate $1/T_1$ shows a $T^{0.5}$ dependence below 3 K and a $T^{0.8}$ dependence in the temperature range of $T$=3.5-14 K (Supplementary Fig. 11), which deviates from the expected $T^2$ behavior. Furthermore, the absence of a significant linear $T$ term in the thermal conductivity implies that the formation of spinon Fermi surface is highly unlikely. Taken together, it seems that the YCu3 system loses the characteristic features of the Dirac spinons in the low-energy and -temperature regime.

To conclude, we have observed the $m$=1/9 magnetization plateau in the Kagomé antiferromagnet YCu$_3$. The underlying Kagomé physics appears to dictate the spin dynamics and field-induced phases of YCu$_3$ possibly due to the limited impact of perturbative terms. The critical field required to access the $m$=1/9 plateau is readily reached by superconducting magnets, rendering YCu$_3$ an unparalleled platform for studying a field-induced quantum entangled phase. This discovery lays the groundwork for deeper exploration and theoretical modeling to gain a comprehensive understanding of the exotic 1/9 plateau state.

**Acknowledgements** The work at SKKU was supported by the National Research Foundation (NRF) of Korea (Grant No. 2020R1A2C3012367 and 2020R1A5A1016518). The work at SNU was financially supported by National Research Foundation of Korea (Grant No. 2019R1A2C22090648) and by the Ministry of Education (Grant No. 2021R1A6C101B418). A portion of this work was performed at the National High Magnetic Field Laboratory, which is supported by the National Science Foundation Cooperative Agreement No. DMR-1644779 and the State of Florida and the U.S. Department of Energy. M. L. is supported by the U.S. Department of Energy, Office of Science, National Quantum Information Science Research Centers. D.W. acknowledges support from the Institute of Basic Science (IBS-R009-Y3). The work at MPK/POSTECH was supported through the National Research Foundation (NRF) funded by MSIP of Korea (Grant No. 2022M3H4A1A04074153).

**Figure Legends/Captions**

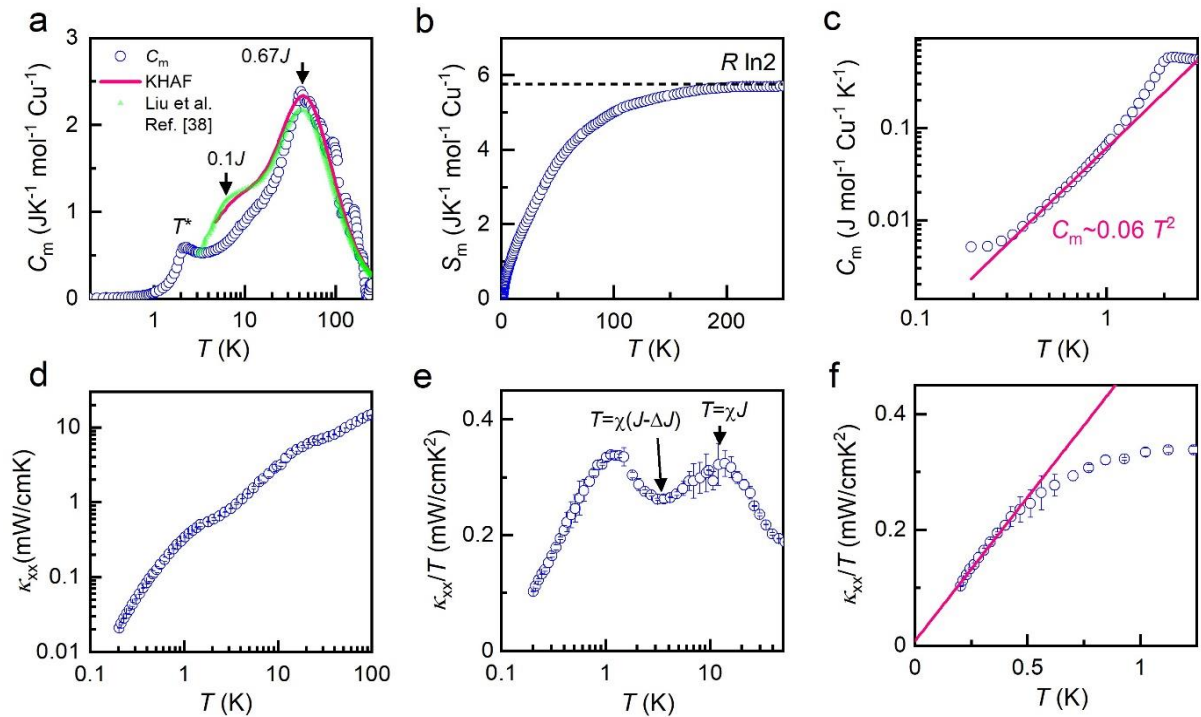

**Figure 1 Magnetic specific heat and thermal conductivity of $YCu_3(OD)_{6.5}Br_{2.5}$.** **a** Semilog plot of the magnetic specific heat ($C_m$; open circles) vs $T$ obtained after subtracting the lattice contribution. The pink line and green triangles represent numerical calculations of $C_m$ for an ideal Kagome antiferromagnet[39]. The two vertical arrows denote two characteristic temperatures at 0.1 $J$ and 0.67 $J$ and $T^*$ represents a weak specific anomaly. **b** Magnetic entropy $S_m$ vs $T$ calculated by integrating $C_m(T)/T$. In the high-temperature limit, the full $R\ln2$ spin entropy (horizontal dashed line) is recovered. **c** The very low-temperature $C_m$ in a log-log scale. The solid line indicates a power-law behavior $C_m \propto T^2$. **d** Plot of thermal conductivity $\kappa$ vs $T$ in the log-log scale. **e** A semi-log plot of $\kappa/T$ vs $T$. **f** Temperature dependence of $\kappa/T$ in the range of $T=0.2 – 1.25$ K. The solid red line represents a linear fit of $\kappa/T$. Thermal conductivity was measured three times at each temperature and the data points were subsequently averaged. The error bars in **d-f** indicate the standard deviation of the data points.

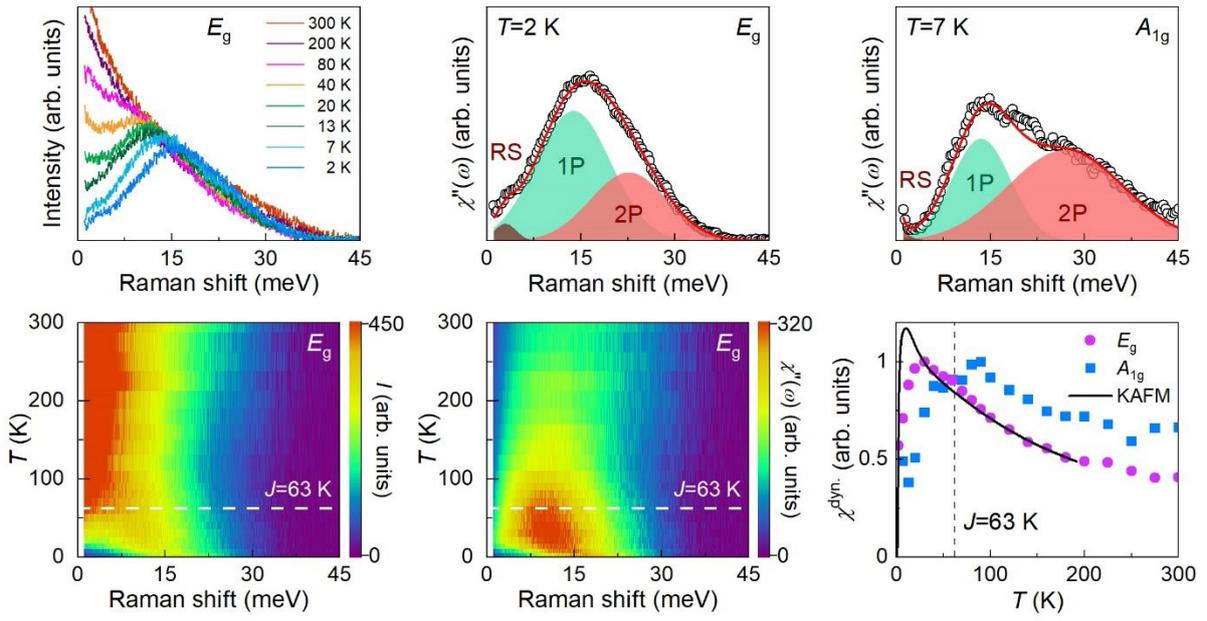

**Fig. 2 Spinon continuum and dynamic Raman susceptibility of YCu$_3$(OD)$_{6.5}$Br$_{2.5}$. a** Temperature dependence of the magnetic Raman spectra obtained after subtracting phonon peaks in the $E_g$ channel. **b,c** Representative Raman susceptibility $\chi''(\omega)$ (open circles) in the $E_g$ channel at $T$=2 K and in the $A_{1g}$ channel at $T$=7 K. The magnetic excitation is decomposed into three components: one-pair (1P; turquoise shading) and two-pair (2P; salmon) spinon-antispinon excitations with additional low-energy excitation (RS; brown). The red solid lines represent the sum of three Gaussian line profiles. **d,e** Color plot of the Raman intensity $I_{E_g}(\omega)$ and $\chi''_{E_g}(\omega, T)$ in the $T$-$\omega$ plane. The horizontal dashed lines indicate the temperature scale of the antiferromagnetic exchange strength $J$=63 K determined from the specific heat. **f** Temperature dependence of the dynamic Raman susceptibilities $\chi^{dyn}_{E_g}(T)$ (pink circles) and $\chi^{dyn}_{A_{1g}}(T)$ (cyan squares) deduced from $\chi''(\omega)s$ through the Kramers-Kronig relation plotted together with the theoretical static magnetic susceptibility (blue solid line) for a perfect Kagomé lattice.

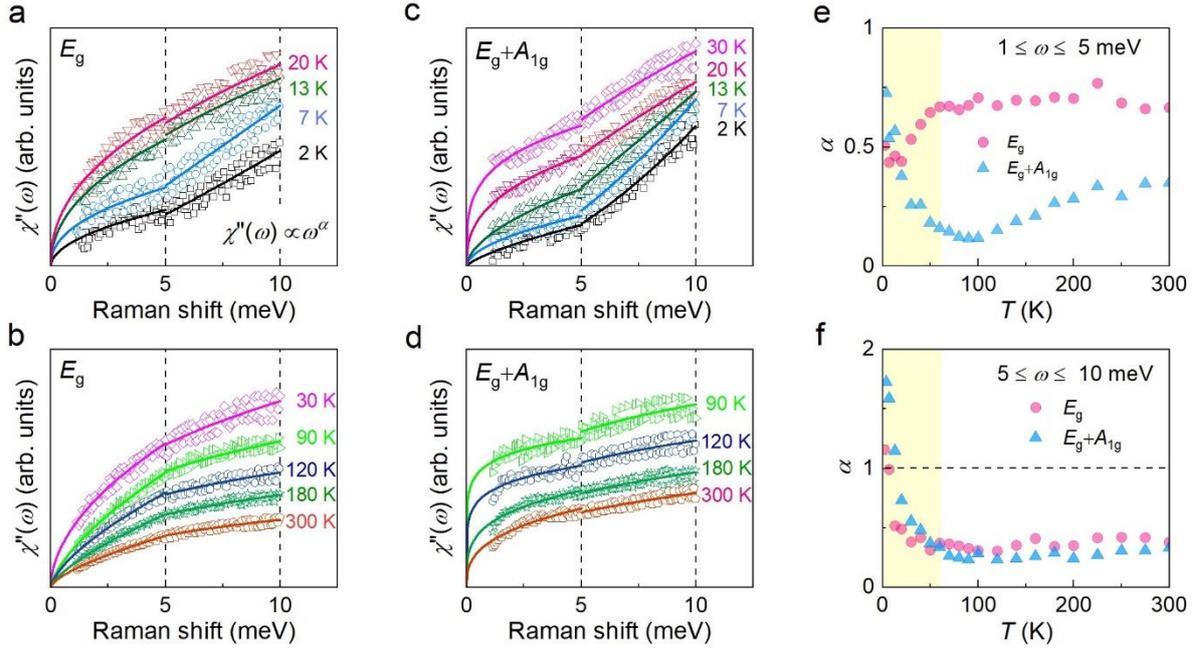

**Fig. 3 Power-law analysis of magnetic Raman susceptibility. a-d** Low-frequency Raman susceptibilities at selected temperatures in the $E_g$ (**a,b**) and $E_g+A_{1g}$ (**c,d**) channels. The vertical dashed lines mark two energy regimes: 1-5 meV and 5-10 meV. The solid lines are power-law fittings to $\chi''(\omega) \propto \omega^\alpha$. **e,f** Temperature dependence of the extracted exponent $\alpha$ in the $E_g$ and $E_g+A_{1g}$ channels and the two different energy windows. The horizontal dashed line denotes $\alpha=1$.

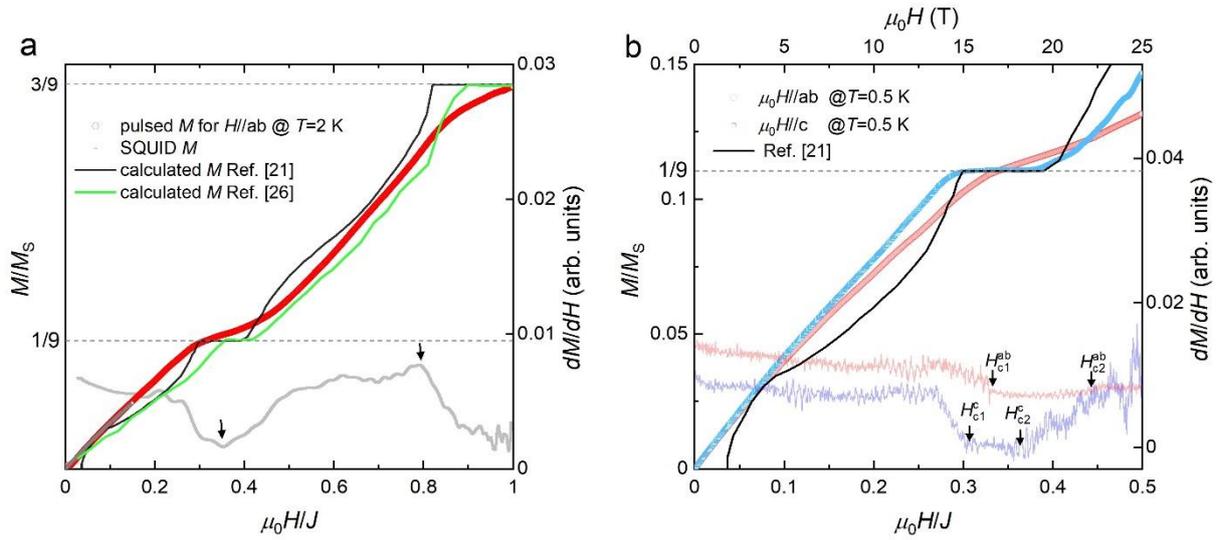

**Fig. 4 High-field magnetization and 1/9 magnetization plateau. a** The pulsed-field magnetization $m(H)$ of YCu$_3$ (magenta circles) and its field derivative d$m$/d$H$ (gray squares) measured at $T$=2 K with a non-destructive pulse magnet up to 60 T. The blue solid line represents the calculated magnetization for the spin-1/2 KHAF, taken from Ref. [26]. The pulsed-field data is calibrated to the low-field SQUID data (yellow triangles). **b** Comparison of the in-plane (orange circles) and out-of-plane (azure squares) magnetization measured at $T$=0.5 K, along with the calculated curve[28] (solid line). The vertical arrows in the d$m$/d$H$ data (violet and coral lines) indicate the lower and upper critical fields of the 1/9 magnetization plateau. The horizontal dashed lines denote the theoretical 1/9 and 3/9 magnetization plateaus.

## Methods

### Crystal growth and magnetic characterization

Single crystals of YCu$_3$(OH)$_{6.5}$Br$_{2.5}$ and deuterated YCu$_3$(OD)$_{6.5}$Br$_{2.5}$ were synthesized by a hydrothermal method. Cu(NO$_3$)$_2$·3H$_2$O (10.0 mmol), Y(NO$_3$)$_3$·6H$_2$O (20.0 mmol), and KBr (60.0 mmol) were mixed in 8.0 mL of distilled water or D$_2$O (7.0 mL). Each of those mixtures was transferred into a 100 mL Teflon-lined stainless-steel autoclave, heated up to 503 K at a rate of 1 K/min, and then maintained for three days. The autoclave was then cooled down to room temperature at a rate of 0.1 K/min. Green hexagonal tablet crystals were isolated by washing with distilled water.

We measured dc magnetization ($\mu_0H$=0 - 7 T) and magnetic susceptibility ($T$=2 – 350 K in an applied magnetic field of $\mu_0H$=1 T) using a superconducting quantum interference device vibrating sample magnetometer (SQUID-VSM, Quantum Design) as presented in Supplementary Note 1.

### Specific heat and thermal conductivity

Thermal conductivity measurements were conducted done by four probes steady-state method using Leiden CF-150 dilution refrigerator (DR) and Quantum Design physical property measurement system (PPMS). Temperature differences were measured using two RuO$_x$ sensors (Rx102A) in the DR or two cernox sensors (Cx1010) in the PPMS. Four 50 μm silver wires were used to make a good thermal link between the sample and thermometers. Heat current was applied parallel to the *ab*-plane of the sample. All the data were measured within the condition of temperature difference $dT$, satisfying $dT/T_0$ <1.5%, where $T_0$ is the base temperature of the specimen.

**High-field magnetization**

Pulsed-field magnetization was measured on high-quality single-crystalline samples using the National High Magnetic Field Laboratory's pulsed-field magnetometry up to 60 T and at the Institute for Materials Research, Tohoku University using a multilayer pulse magnet up to 30 T. The pickup coil is a radially compensated coil wound from 50 American wire gauge copper wire. The samples were inserted into a nonmagnetic ampule and fixed with vacuum grease. The voltage signal induced in the coil in pulsed fields is proportional to the derivative of magnetization with time. The voltage is numerically integrated and calibrated with the magnetization obtained in dc magnetization measured using SQUID-VSM (Quantum Design). For each field-dependent magnetization curve, two data with the sample-in and out of the coil signals under identical conditions were collected, and the sample-out signal was subtracted from the sample-in signal to remove background signals.

**Raman scattering measurements**

Raman scattering measurements were carried out in exact backscattering geometry using the excitation line λ=561 nm laser (Oxxius LCX). We collected the scattered spectra with a TriVista spectrometer (Princeton Instruments, SP-2500i) equipped with a liquid-Nitrogen cooled CCD (PyLoN eXcellon). We used a volume Bragg grating notch filter (Optigrate), which enabled low-frequency Raman measurements down to ~10 cm$^{-1}$. The incident laser was focused on a spot diameter of about 4 μm of the sample surface with a ×40 microscope objective. The laser power was kept below $P$=0.2 mW to minimize heating effects. The samples were mounted onto a liquid helium flow cryostat while varying the temperature between $T$=2 K and 300 K.

**Data availability**

All relevant data supporting the findings of this study are available from the corresponding authors on reasonable request.

# Supplementary Information

# Table of Contents

## Supplementary Section 1. Magnetic susceptibility and specific heat of YCu₃(OD)₆.₅Br₂.₅ and YCu₃(OH)₆.₅Br₂.₅

Supplementary Fig. 1a exhibits the temperature dependence of dc magnetic susceptibilities $\chi(T)$ of YCu₃(OD)₆.₅Br₂.₅ and YCu₃(OH)₆.₅Br₂.₅ for $H//ab$ and $H\perp ab$. With decreasing temperature toward $T=0$ K, $\chi(T)$ shows a steep increase, which is attributed to bond randomness or Dzyaloshinskii-Moriya interaction born from the bond randomness. As shown in Supplementary Fig. 1b,c, the inverse magnetic susceptibility $1/\chi$ follows the Curie–Weiss law in the high-temperature range of $T=100$ -350 K. The extracted Curie-Weiss temperatures are $\Theta_{CW}^{\parallel} = -88.36(2)$ K and $\Theta_{CW}^{\perp} = -99.88(2)$ K, and the effective magnetic moments are $\mu_{eff}^{\parallel}=1.90(1)$ $\mu_B$ and $\mu_{eff}^{\perp}=2.07(0)$ $\mu_B$ for YCu₃(OD)₆.₅Br₂.₅ and $\Theta_{CW}^{\parallel}= -104.07(1)$ K and $\Theta_{CW}^{\perp}=-117.57(0)$ K and the effective magnetic moment $\mu_{eff}^{\parallel}=1.95(0)$ $\mu_B$ and $\mu_{eff}^{\perp}=2.11(6)$ $\mu_B$ for YCu₃(OH)₆.₅Br₂.₅.

In Supplementary Fig. 1d, we plot the temperature dependence of the specific heat $C_p(T)$. To single out the magnetic contribution $C_m(T)$ from $C_p(T)$, we calculate the lattice contribution $C_{lat}(T)$ based on the combination of one Debye and three Einstein functions as

$$C_{lat}(T) = 9C_D N k_B \left(\frac{T}{\theta_D}\right)^3 \int_0^{\theta_D/T} \frac{x^4 e^x}{(e^x-1)^2} dx$$

$$+ 3Nk_B \sum_{i=1-3} C_E^i \left(\frac{\theta_E^i}{T}\right)^2 \frac{e^{(\theta_E^i/T)}}{\left(e^{(\theta_E^i/T)}-1\right)^2}.$$

Here, $C_D$ and $C_E^i$ are the weighting factors and $\theta_D$ and $\theta_E^i$ are the Debye and the Einstein temperature, respectively. From the fittings above 100 K, we obtain $C_D = 1.390$, $C_E^1 = 1.000$, $C_E^2 = 2.800$, $C_E^3 = 1.310$, $\theta_D = 151.0$ K, $\theta_E^1 = 255.0$ K, $\theta_E^2 = 636.0$ K, and $\theta_E^3 = 1641.0$ K. The calculated $C_{lat}(T)$ is denoted by the solid pink line. The obtained $C_m(T)= C_p(T)-C_{lat}(T)$ is plotted in Fig. 1a of the main text and is used to evaluate the magnetic entropy.

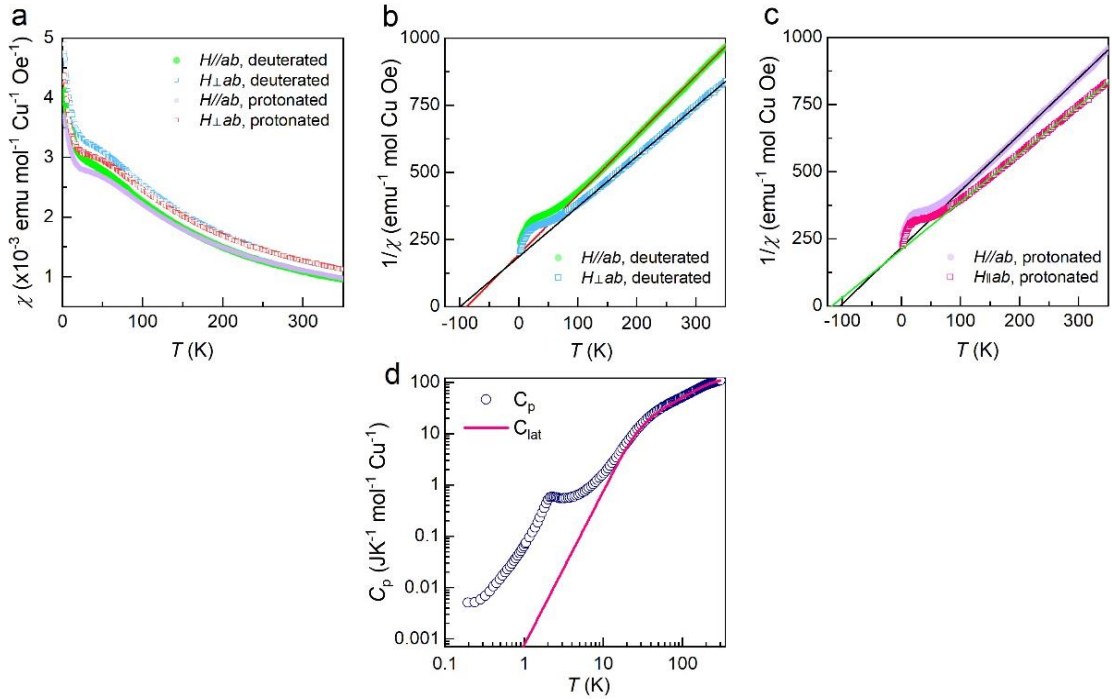

**Fig. S1 Magnetic susceptibility and heat capacity. a** Temperature dependence of the static magnetic susceptibility of YCu$_3$(OD)$_{6.5}$Br$_{2.5}$ (green and light blue symbols) and YCu$_3$(OH)$_{6.5}$Br$_{2.5}$ (red and purple symbols) for $H//ab$ and $H\perp ab$. **b,c** Inverse magnetic susceptibility of YCu$_3$(OD)$_{6.5}$Br$_{2.5}$ and YCu$_3$(OH)$_{6.5}$Br$_{2.5}$ as a function of temperature. The solid lines are fittings to the Curie-Weiss law. **d** The open blue circles represent the measured magnetic heat capacity. The solid magenta curve denotes the evaluated lattice contributions using one Debye model and two Einstein terms.

## Supplementary Section 2. Temperature dependence of Raman spectra of YCu$_3$(OD)$_{6.5}$Br$_{2.5}$

Supplementary Fig. 2 present the Raman intensity $I(\omega)$ in (*XY*) and (*XX*) polarizations measured at selected temperatures from 300 K to base temperature. According to the point group representation of *D3d* (*-3m*) [1], the factor-group analysis of YCu$_3$(OD)$_{6.5}$Br$_{2.5}$ with space group *P-3m1* predicts the nine Raman-active phonons of $\Gamma = 4A_{1g}(aa,bb,cc) + 5E_g(aa,bb,ab,bc,ca)$. We observe a total of $7A_{1g}$ and $6E_g$ modes for energies below 50 meV. The extra modes, often weak in intensity or appearing as shoulders of other phonons, are ascribed due to local distortions induced by the random occupation of the OH$^-$ and Br$^-$ anions. In addition, a magnetic continuum superimposes the phonon modes, which experience a systematic thermal evolution. The phonon-subtracted magnetic excitations are discussed in the main text.

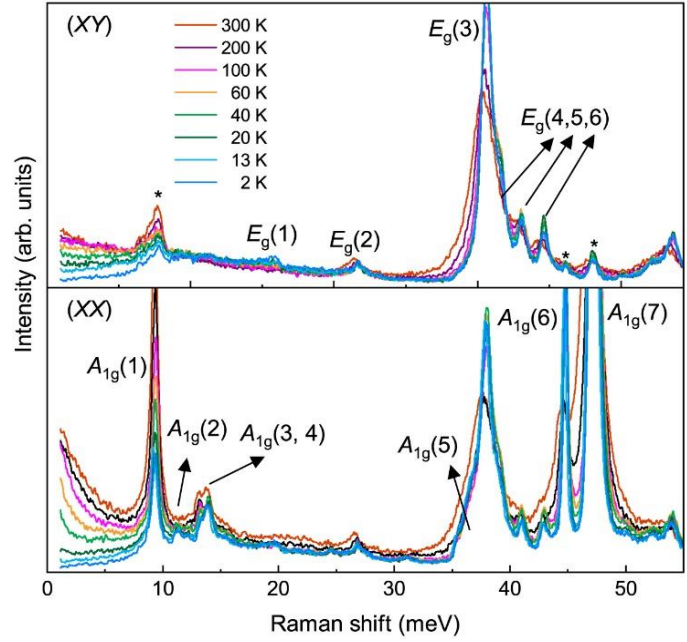

**Fig. S2 As-measured Raman spectra of $YCu_3(OD)_{6.5}Br_{2.5}$.** Temperature dependence of Raman spectra measured in the temperature range of $T$=2 - 300 K in ($XY$) and ($XX$) polarizations. The asterisk symbols mark the phonon modes that show up due to polarizer leakage.

In Supplementary Fig. 3, we detail the phonon parameters: the frequency $\omega(T)$, the full width at half maximum $\Gamma(T)$, and the normalized intensity for the representative $A_{1g}$ phonon modes at 9.34 meV and 47.2 meV. For a quantitative analysis of the phonon parameters, we fit the observed data to a simple second-order anharmonic phonon decay model [2]:

$$\omega(T) = \omega_0 + A[1 + 2/(e^{\hbar\omega_0/2k_BT} - 1)],$$

and
$$\Gamma(T) = \Gamma_0 + B[1 + 2/(e^{\hbar\omega_0/2k_BT} - 1)].$$

Here, $\omega_0$ and $\Gamma_0$ are the phonon frequency and linewidth at $T$=0 K, and $A$ and $B$ are constants. The high-energy phonon mode shows a strong deviation from the fitted curves with parameters $A$=-0.03 meV and $B$=0.054 meV for temperatures below $T$=150 K. The low-energy phonon mode agrees well in the temperature range of $T$=70−300 K with the fitting parameters $A$=-9.87×10$^{-4}$ meV and $B$=8.19×10$^{-4}$ meV. Overall, the renormalization of the phonon energy and the additional relaxation channel point toward the presence of spin-lattice coupling.

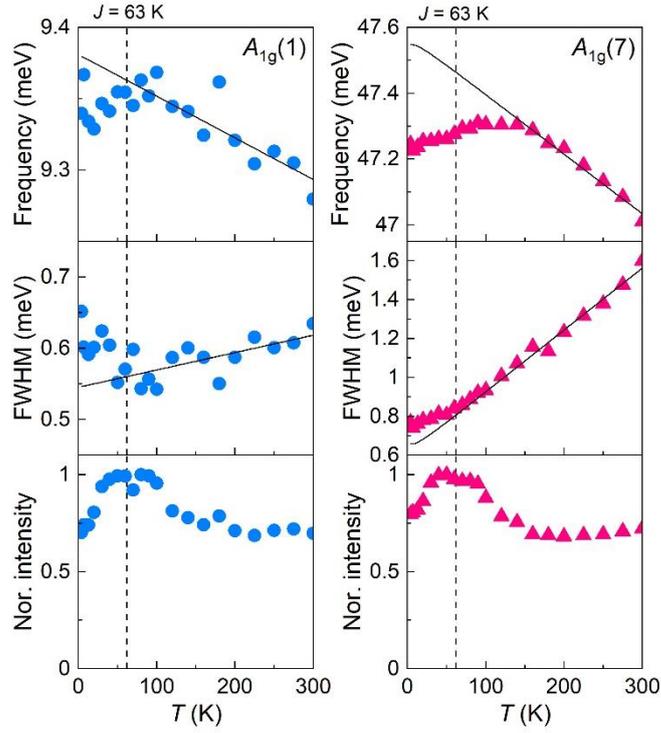

**Fig. S3 Temperature dependence of the phonon parameters of $YCu_3(OD)_{6.5}Br_{2.5}$.** Temperature dependence of the frequency, the FWHM, and the normalized intensity of the two $A_g$ modes. The dashed vertical line marks the superexchange energy scale of $J$=63 K. The solid lines are fitted curves using the anharmonic model.

**Supplementary Section 3. Angular dependence of the phonon parameters**

We recorded the angular dependence of the $E_g(3)$ and $A_{1g}(7)$ phonon modes. In the cross (*XY*) and parallel (*XX*) polarization, a superachromatic l/2 waveplate placed above the focusing microscope objective rotates the light polarization of the incident and scattered light in steps of 8° between 0° and 360°. Supplementary Fig. 4 summarizes the angular dependence of the two phonon and spinon excitations.

According to the factor group analysis, we can observe the $E_g(3)$ and $A_{1g}(7)$ modes in the (*ab*) plane. The Raman intensity of the phonon modes is given as [3]

$$I \propto |e_{in} \cdot R \cdot e_{out}|^2, \tag{1}$$

where $e_{in}$ and $e_{out}$ are the polarization vectors of the incident and scattered light, respectively. The Raman tensors are given as

$$R(E_g) = \begin{pmatrix} c & 0 & 0 \\ 0 & -c & d \\ 0 & d & 0 \end{pmatrix}, \begin{pmatrix} 0 & -c & -d \\ -c & 0 & 0 \\ -d & 0 & 0 \end{pmatrix} \tag{2}$$

and

$$R(A_{1g}) = \begin{pmatrix} a & 0 & 0 \\ 0 & a & 0 \\ 0 & 0 & b \end{pmatrix}. \tag{3}$$

The polarization unit vectors are expressed as $e_{in} = (\cos\theta, \sin\theta, 0)$ and $e_{out} = (-\sin\theta, \cos\theta, 0)$ for (XY) polarization. Then, we obtain the angular dependence of the Raman intensity for the $A_{1g}$ and $E_g$ modes:

$$I(E_g) = |c|^2, \tag{4}$$

and

$$I(A_{1g}) = 0. \tag{5}$$

In (XY) polarization, the intensity of the $E_g$ mode is independent of an angle. The $A_{1g}$ mode vanishes when the incident and scattered polarization are perpendicular to each other.

For (XX) polarization, we have $e_{in} = e_{out} = (\cos\theta, \sin\theta, 0)$. The Raman intensity for each mode is given by

$$I(E_g) = |c|^2, \tag{6}$$

and

$$I(A_{1g}) = |a|^2. \tag{7}$$

The intensities of both modes are independent of an angle.

We compare the calculated intensities to the experimental angular-dependent data. Supplementary Fig. 4a,b show polar plots of the angular dependence of the $E_g$ (3) (ω=37.95 meV) and $A_{1g}$(7) (ω=47.2 meV). The data are in good agreement with the theoretical curves.

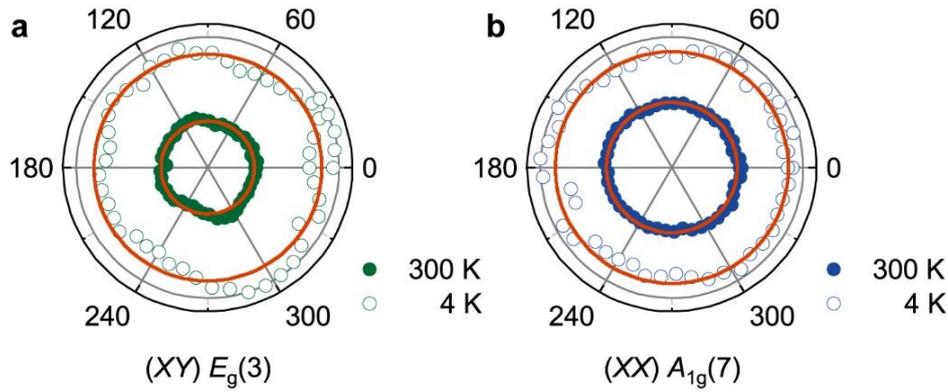

**Fig. S4 Angular dependence of the phonon modes. a,b** Angular dependence of the $E_g(3)$ and $A_{1g}(7)$ phonon modes for ($XY$) and ($XX$) polarizations at $T$=300 K and 4 K.

**Supplementary Section 4. Comparison of magnetic excitations between YCu$_3$(OD)$_{6.5}$Br$_{2.5}$ and its magnetic counterpart YCu$_3$(OH)$_6$Cl$_3$ and their angular dependence**

To ascertain the spinon nature of low-lying spin excitations, we compare the $E_g$ Raman spectra between YCu$_3$(OD)$_{6.5}$Br$_{2.5}$ (green line) and its magnetically ordered counterpart YCu$_3$(OH)$_6$Cl$_3$ (navy line) with $T_N$=15 K at $T$=2 K in Supplementary Fig. 5a. Further, we plot the magnetic excitations of Cu$_3$Zn(OH)$_6$FBr (black line) and its magnetically ordered sister EuCu$_3$(OH)$_6$Cl$_3$ with $T_N$=17 K (red line) [4]. To ensure consistent comparison, the exchange interaction of Zn-barlowite is adequately scaled.

Remarkably, the magnetically ordered compounds YCu$_3$(OH)$_6$Cl$_3$ and EuCu$_3$(OH)$_6$Cl$_3$ commonly bear a sharp peak, which is assigned to one-magnon scattering whose energy scale is comparable to that of the one-pair spinon-antispinon continuum. Supplementary Fig. 5b presents the temperature dependence of magnetic excitations for YCu$_3$(OH)$_6$Cl$_3$. With increasing temperature toward $T_N$=15 K, the one-magnon excitation is systematically suppressed, supporting the one-magnon interpretation. In sharp contrast, YCu$_3$(OD)$_{6.5}$Br$_{2.5}$ and Cu$_3$Zn(OH)$_6$FBr feature a broad magnetic continuum. The absence of magnon excitations confirms that the magnetic continuum of YCu$_3$(OD)$_{6.5}$Br$_{2.5}$ is governed by spinon excitations.

In Supplementary Fig. 6, we present the color plots of the phonon-subtracted Raman spectra in ($XY$) and ($XX$) polarizations at $T$= 300 K and 4 K. Supplementary Fig. 6c and f show the angular dependence of intensities integrated up to 45 meV. We note that the magnetic Raman excitation exhibits no angle-dependent behaviors as the phonons do.

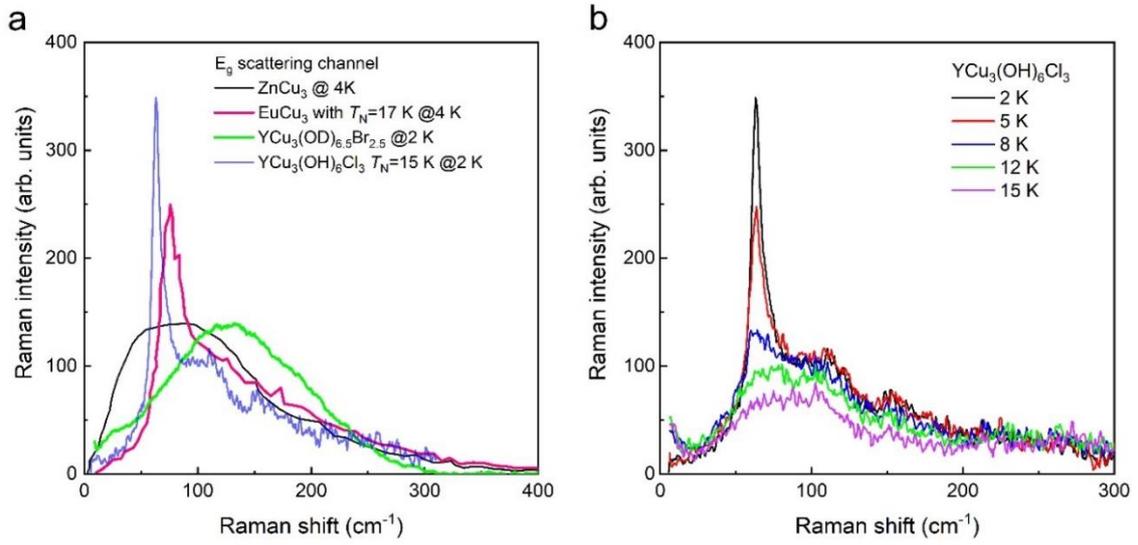

**Fig. S5 Comparison of $E_g$ magnetic Raman spectra of Kagome materials.** **a** Magnetic Raman spectrum of four Kagome candidate materials measured at $T$=2-4 K in the $E_g$ scattering channel. **b** Thermal evolution of magnetic excitations of YCu$_3$(OH)$_6$Cl$_3$ in the antiferromagnetically ordered state below $T_N$=15 K.

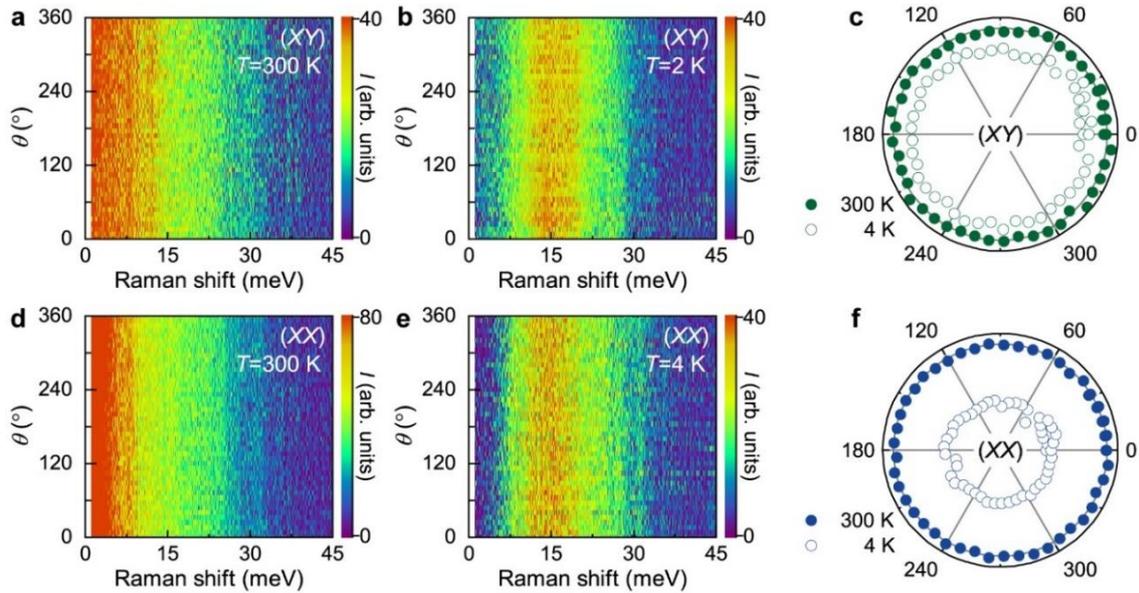

**Fig. S6 Angular dependence of spinon excitations.** **a-c** Angular dependence of spinon excitations at $T$=300 K and 4 K and the angular dependence of their intensity integrated up to 45 meV for ($XY$) polarization. **d-f** Angular dependence of spinon excitations at $T$=300 K and 4 K and the angular dependence of the intensity integrated up to 45 meV for ($XX$) polarization.

**Supplementary Section 5. Modeling a magnetic continuum**

To crosscheck whether a magnetic continuum mainly comprises one-pair (1P) and two-pair (2P) spinon-antispinon excitations, we analyze its temperature dependence in two different models. Specifically, we decompose the magnetic Raman spectra into three and four components.

As plotted in Supplementary Fig. 7a, the three-component model consists of 1P, 2P, and RS excitations. The 1P component involves largely one-pair spinon-antispinon excitations, while the 2P component represents two-pair spinon-antispinon excitations. The RS component originates mainly from random-singlet-like (RS) excitations. Each contribution is approximated by Gaussian line profiles. Supplementary Fig. 7b plots the temperature dependence of the peak energy and FWHM of each component. With increasing temperature through 13 K, the RS excitation evolves into quasielastic scattering (QES). On the other hand, on heating to 20 K, the 1P and 2P excitations soften rapidly and then undergo a gradual renormalization of their energy. In parallel, the 1P and 2P excitations experience substantial thermal damping up to 20 K, and their FWHMs become independent of temperature above 20 K. The temperature scale of 20 K is associated with a coherent-to-incoherent crossover of spinons, inferred from the thermal conductivity data in Fig. 1e of the main text. The systematic thermal variation of each component gives credence to the validity of the employed model.

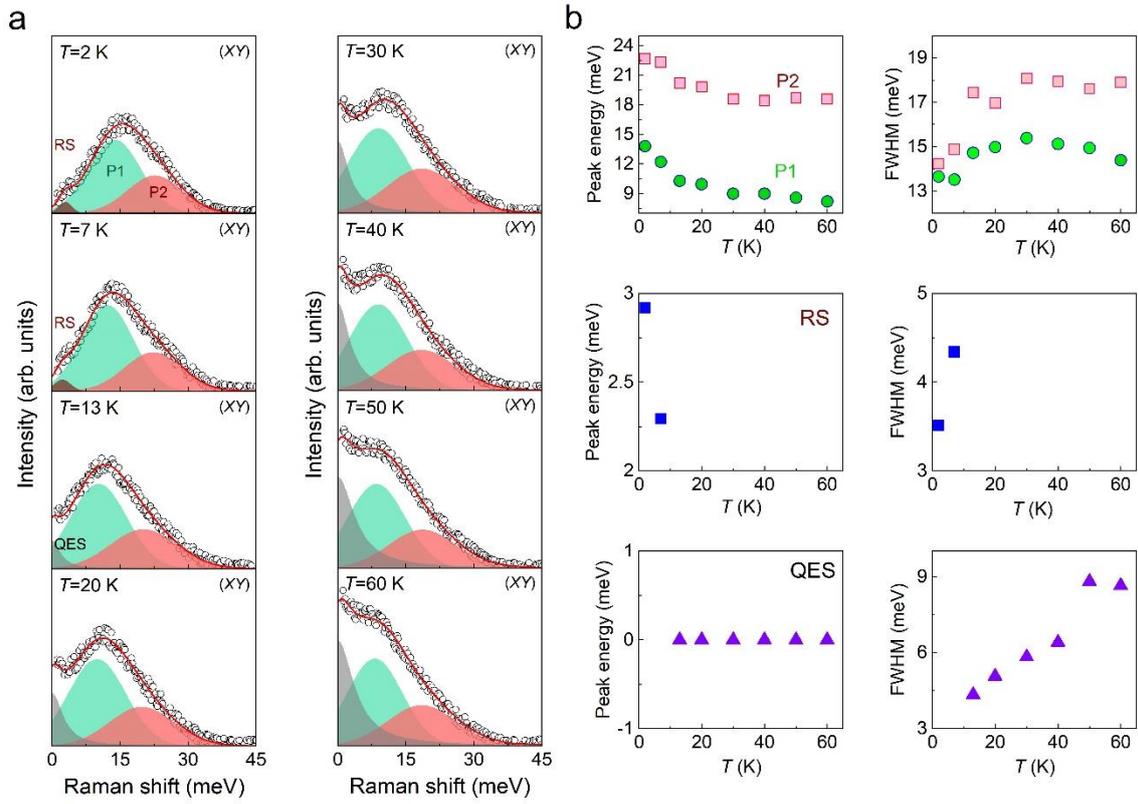

**Fig. S7 Temperature dependence of spinon excitations in the three-component decomposition model. a** Thermal variation of one-pair and two-pair spinon-antispinon excitations together with random-singlet-like excitation. **b** Temperature dependence of the peak energy and the full width at half maximum of the magnetic excitations.

We repeated the same analysis based on the four-component model. The fitting results are summarized in Supplementary Fig. 8. Overall, we find that the 1P component gains a spectral weight relative to the 2P and 3P components. Its high-energy cutoff extends to 30 meV, much larger than the one-spinon bandwidth of 13 meV (=$6\chi J$ with $J$=63 K and $\chi$=0.4). Further, the FWHM of the 2P and 3P excitations is way smaller than that of the 1P excitation. Thus, in this model, the 2P and 3P excitations cannot be regarded as multi-spinon excitations.

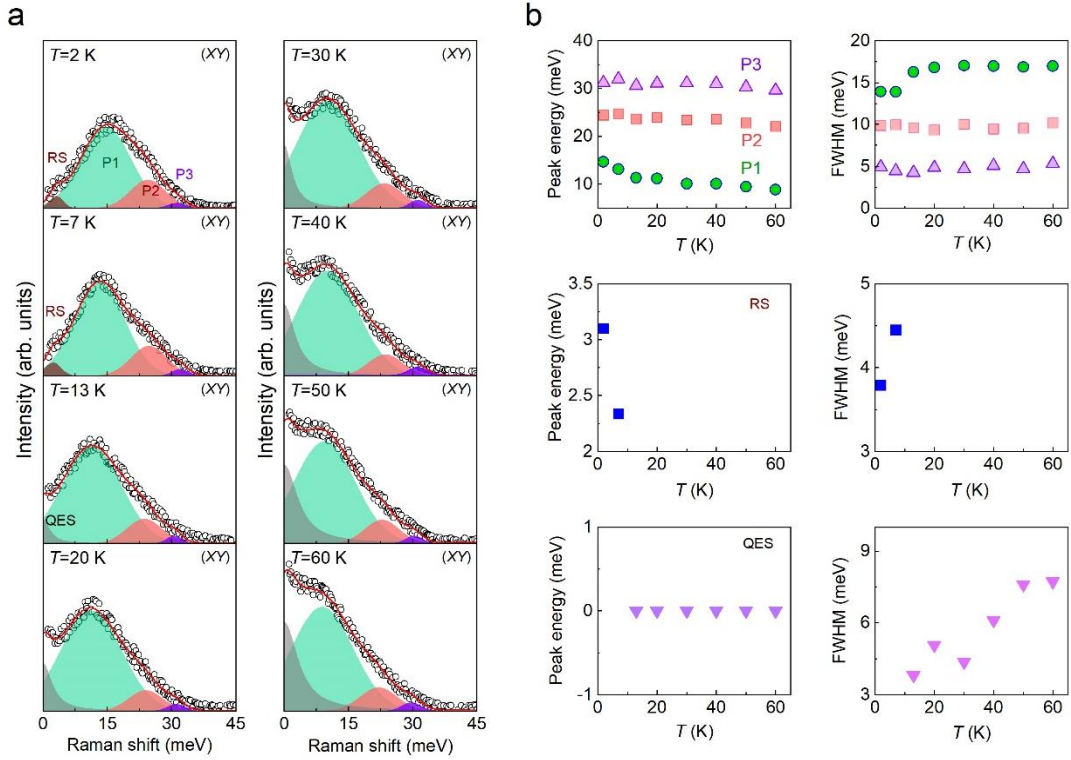

**Fig. S8 Temperature dependence of spinon excitations in the four-component decomposition model. a** Thermal evolution of the P1, P2, P3, and RS excitations. **b** Temperature dependence of the peak energy and the full width at half maximum of the P1, P2, P3, and RS magnetic excitations.

**Supplementary Section 6. Temperature dependence of Raman susceptibility $\chi''_{XX}$ and $\chi''_{A_{1g}}$**

In a spin-$\frac{1}{2}$ Kagomé lattice, the magnetic Raman spectrum comprises spinon excitations in all symmetry channels. To study the symmetry dependence of spinon excitations, we measured Raman scattering in (*XY*) and (*XX*) polarizations. In Supplementary Fig. 9, we summarize the thermal evolution of the Raman intensity, the Raman susceptibility $\chi''(\omega)$, the dynamic Raman susceptibility $\chi^{\text{dyn}}_{XX}$, and the peak energies $\omega_{1P}(T)$ and $\omega_{2P}(T)$. In (*XX*) polarization, the $E_g$ and $A_{1g}$ contributions are intermingled. As such, we first single out the pure $A_{1g}$ Raman susceptibility $\chi''_{A_{1g}} = \chi''_{XX} - \chi''_{XY}$ (Supplementary Fig. 10).

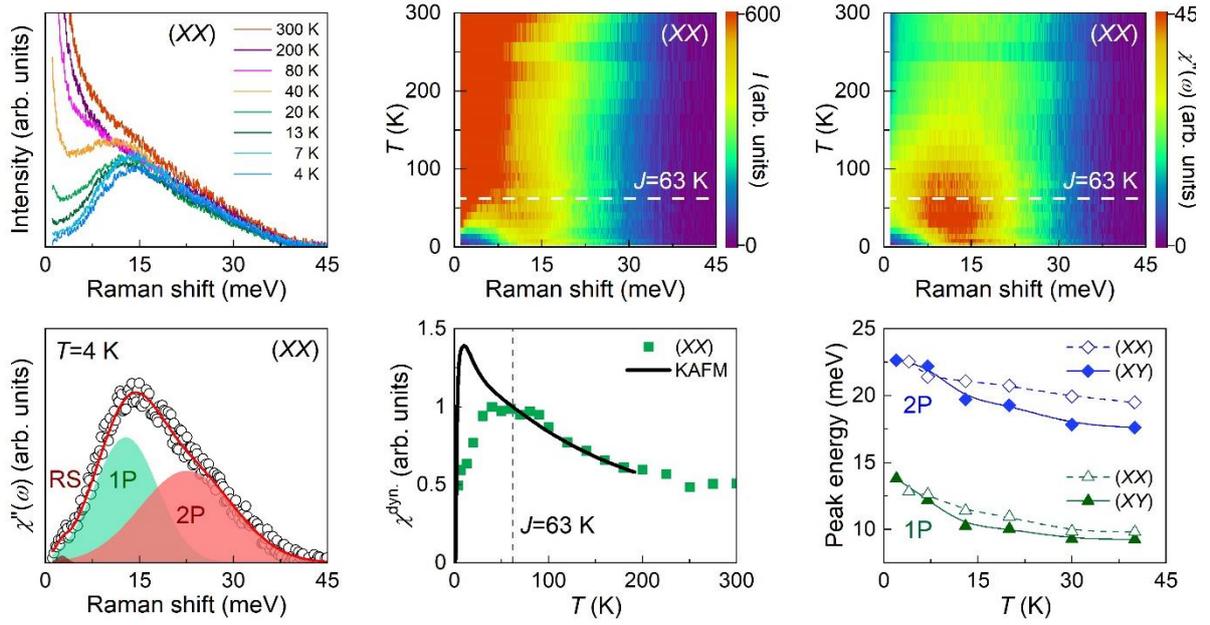

**Fig. S9 Spinon excitations and dynamic susceptibility in (*XX*) polarization. a** Temperature dependence of the magnetic excitations in (*XX*) polarization. **b,c** Color plots of the Raman intensity $I(\omega)$ and the Raman susceptibility $\chi''(\omega)$ in the *T*-$\omega$ plane. **d** The *T*=4 K Raman susceptibility $\chi''(\omega)$ comprises two spinon excitations: one-pair spinon (1P) and two-pair spinon (2P) along with a random-singlet state (RS). **e** Temperature dependence of the dynamic Raman susceptibility derived from $\chi^{\mathrm{dyn}} = \lim_{\omega\to 0}\chi(\mathbf{k}=0,\omega) \equiv \frac{2}{\pi}\int_0^{45\,\mathrm{meV}}\frac{\chi''(\omega)}{\omega}d\omega$. The solid black line represents the static magnetic susceptibility calculated for an ideal Kagomé antiferromagnet taken from Ref. [5]. **f** Peak energy of the 1P and 2P spinon excitations as a function of temperature for (*XX*) and (*XY*) polarizations.

As shown in Supplementary Fig. 9a-d, the magnetic continuum in (*XX*) polarization mainly consists of one-pair (1P) and two-pair (2P) spinon-antispinon excitations. Compared to the $E_g$ channel, the spectral weight of the spinon excitations becomes smaller in the $A_{1g}$ channel. Additionally, the ratio of the 2P to the 1P spectral weight $\frac{\mathrm{Area}(2P)}{\mathrm{Area}(1P)}$=1.85 for the $A_{1g}$ channel is reduced to 0.57 for the $E_g$ channel. This suggests that the 1P excitation dominates in the $E_g$ channel. Conversely, the 2P excitation is dominant in the $A_{1g}$ channel.

We next turn to the $A_{1g}$ Raman susceptibility $\chi''_{A_{1g}}$. In contrast to $\chi''_{E_g}$, $\chi''_{A_{1g}}$ shows a quick melting of the spinon excitations into a quasielastic response above *T*=30 K as shown in

Supplementary Fig. 10. The peak energies $\omega_{1P}$ and $\omega_{2P}$ soften monotonically with temperature increasing. The dynamic susceptibility $\chi_{A_{1g}}^{\text{dyn}}(T)$ derived from $\chi_{A_{1g}}^{"}(T)$ deviates from the static magnetic susceptibility calculated for a perfect Kagomé antiferromagnet below 80 K.

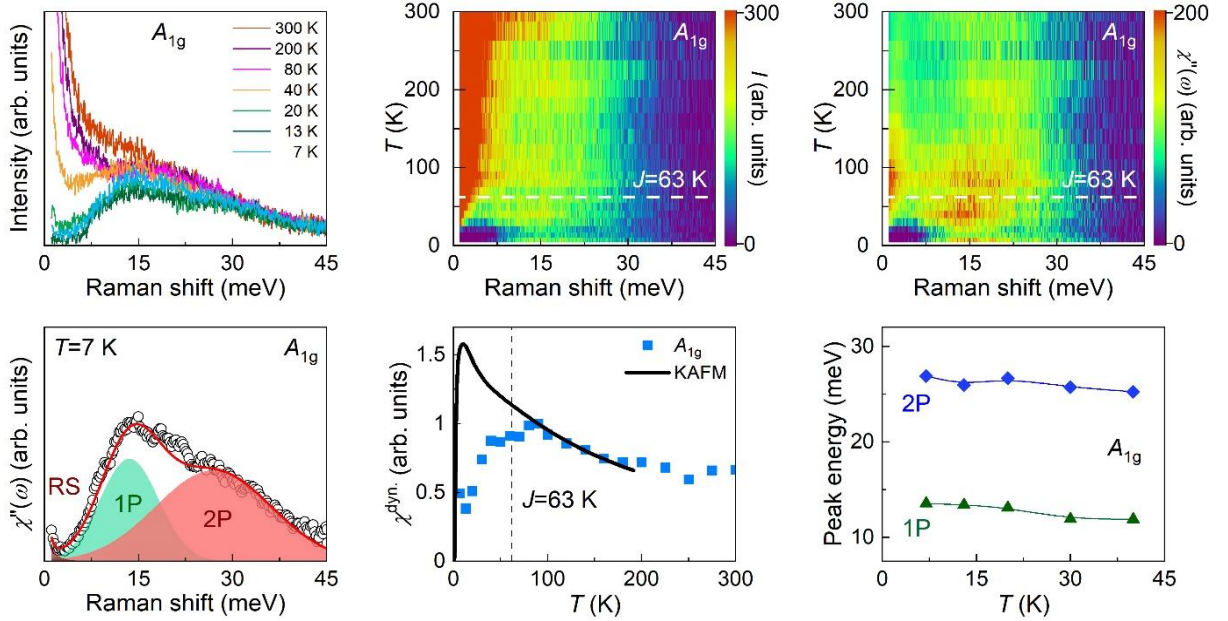

**Fig. S10 Spinon excitations and their dynamic behavior in the $A_{1g}$ channel ($\chi_{A_{1g}}^{"} = \chi_{XX}^{"} - \chi_{XY}^{"}$). a** Temperature dependence of the $A_{1g}$ magnetic excitations. **b,c** Color plots of the Raman intensity $I_{A_{1g}}(\omega)$ and Raman susceptibility $\chi_{A_{1g}}^{"}$ in the $T$-$\omega$ plane. **d** The Raman susceptibility $\chi_{A_{1g}}^{"}$ is decomposed into one-pair spinon (1P) and two-pair spinon (2P) together with a random-singlet state (RS). **e** Temperature dependence of the dynamic Raman susceptibility derived from $\chi^{\text{dyn}} \equiv \frac{2}{\pi}\int_0^{45\,\text{meV}} \frac{\chi^{"}(\omega)}{\omega} d\omega$. The solid black line represents the theoretical static susceptibility for an isotropic kagomé antiferromagnet. **f** Peak energy of the 1P and 2P spinon excitations as a function of temperature.

**Supplementary Section 7. $^2$D NMR spin-lattice relaxation rate**

Supplementary Figure 11 presents the temperature dependence of the spin-lattice relaxation rate $1/T_1$. Upon cooling below $T$=35 K, the $^2$D $1/T_1$ exhibits a gradual decrease, a common characteristic observed in spin-liquid materials. We find that $1/T_1$ follows a power-law behavior $1/T_1 \sim T^\alpha$ with the exponent α=0.8 in the temperature range of $T$=3.5-14 K and the exponent

α=0.5 for temperatures below 3.5 K. These power-law dependencies differ from the expected $T^2$ behavior typically associated with a Dirac spin liquid.

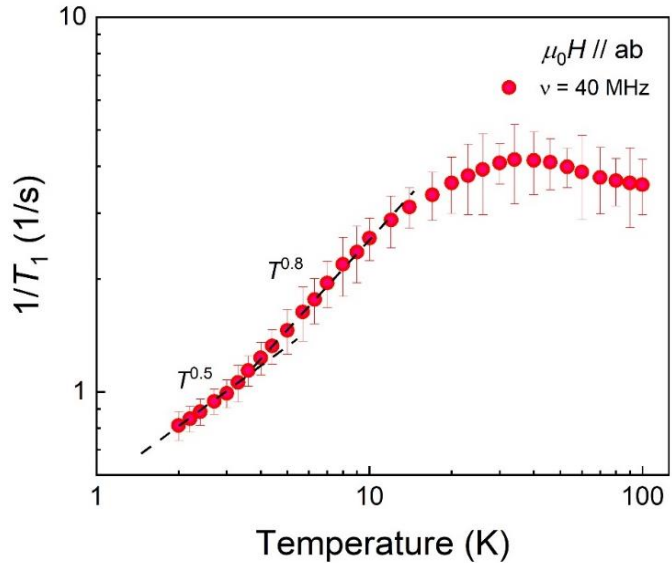

**Fig. S11 Temperature dependence of the spin-lattice relaxation rate $1/T_1$.** The dashed lines are fits to the power-law behavior $1/T_1 \sim T^{0.8}$ in the temperature range of $T$=3.5-14 K and $T^{0.5}$ for temperatures below 3.5 K.

**Supplementary References**